# Towards Semantic Detection of Smells in Cloud Infrastructure Code

Indika Kumara
i.p.k.weerasingha.dewage@tue.nl
Jheronimus Academy of Data Science, Eindhoven
University of Technology, Netherlands

Zoe Vasileiou
Georgios Meditskos
{zvasilei,gmeditsk}@iti.gr
Information Technologies Institute, Centre for Research &
Technology - Hellas, Greece

Damian A. Tamburri
Willem-Jan Van Den Heuvel
{d.a.tamburri,W.J.A.M.vdnHeuvel}@tue.nl
Jheronimus Academy of Data Science, Eindhoven
University of Technology, Netherlands

Anastasios Karakostas
Stefanos Vrochidis
Ioannis Kompatsiaris
{akarakos,stefanos,ikom}@iti.gr
Information Technologies Institute, Centre for Research &
Technology - Hellas, Greece

**Abstract**

Automated deployment and management of Cloud applications relies on descriptions of their deployment topologies, often referred to as Infrastructure Code. As the complexity of applications and their deployment models increases, developers inadvertently introduce software smells to such code specifications, for instance, violations of good coding practices, modular structure, and more. This paper presents a knowledge-driven approach enabling developers to identify the aforementioned smells in deployment descriptions. We detect smells with SPARQL-based rules over pattern-based OWL 2 knowledge graphs capturing deployment models. We show the feasibility of our approach with a prototype and three case studies.

## 1 Introduction

As Cloud computing technologies continue to become mature, organizations are increasingly using Cloud as their IT infrastructure. According to recent Gartner surveys [7], more than a third of organizations consider the adoption of Cloud as a top three priority. Organizations have complex applications, consisting of multiple components that need to be deployed over one or more cloud infrastructures [4, 6]. Thus, automated deployment and management of cloud applications is vitally important.

In recent years, several infrastructure automation tools have been introduced to simplify and automate application deployment, for example, CloudFormation, TerraForm, Puppet, Chef, and Docker Stack. The provisioning processes in most of these tools use an explicit or implicit model of the deployment topology of the application in terms of components and their relationships, and nodes that host the components [2, 6]. The design, specification, and enactment of deployment models has been a key research topic [1, 2, 6]. As the size and complexity of the deployment model increase, it is critical to maintain their quality. To this end, the software smells in the deployment models can be identi- fied and removed. A software smell is any characteristic in the artifacts of the software that possibly indicates a deeper problem or quality issue [16], for example, occurrences of antipatterns, and use of insecure coding practices such as hard-coded secrets and empty passwords [11]. The smells can negatively impact software quality attributes such as maintainability, change proneness, and security [16].

In software engineering research, smell detection has been a key topic [16]. Several studies have used a rule-based approach to detect smells in different artifacts such as object-oriented programs [9], service descriptions [10], and infrastructure automation scripts [11, 15]. The rule-based approach is also popular in industry, for instance, so-called *Lint* tools for Docker, Chef, TerraForm, and Puppet, However, mostly, these tools use informal rules, and operate directly on source code. On the other hand, some studies have employed successfully semantic technologies to specify and detect antipatterns, for example, in software projects [14] and service APIs [3]. In most cases, non-standard rule languages are used (e.g. SWRL[1]), while the underlying ontologies follow specifically-designed conceptual models. To the best of our knowledge, there does not exist any study on semantic approaches to predict smells in deployment model descriptions.

In this paper, we present a semantic approach to detecting smells in deployment model descriptions. We develop the semantic models (ontology) to formally describe a deployment model, reusing the Description and Situation (DnS) pattern [5] implemented in DOLCE ontology. As the different languages are used to specify deployment models, our ontologies are based on a widely used open standard, namely TOSCA (Topology and Orchestration Specification for Cloud Applications)[1, 8]. TOSCA enables standardized descriptions of heterogeneous (e.g., Cloud, Edge, and HPC) distributed applications. We develop SPARQL-based rules over our ontologies to detect deployment model smells.

The rest of the paper is organized as follows. Section 2 provides an overview of TOSCA, and summarizes the related work. Section 3 presents our approach in detail, including ontologies and smell detection rules. Section 4 describes the prototype implementation and the evaluation of our approach. Section 5 concludes the paper.

## 2 Background and Related Work

### 2.1 TOSCA Overview

TOSCA [1, 8] is an OASIS standard for describing deployment and management of Cloud applications. The key TOSCA concepts for describing a deployment model are : *Topology Tem- plate*, *Node Template*, *Node Type*, *Relationship Template*, and *Relationship Type*. *Topology Template* specifies the structure of the application in terms of *Node Templates* and *Relationship Templates*. Node Templates model application components (e.g., virtual machines, databases, and web services), whose semantics (e.g., properties, attributes, requirements, capabili- ties and interfaces) are defined by *Node Types*. *Relationship templates* capture relations between the nodes, for example, a node hosting another node or network connection between nodes. *Relationship types* specify the semantics (e.g., proper- ties and interfaces) of these relationships. The properties and attributes represent the desired and actual states of nodes or relationships, e.g., IP address or VM image type. Interfaces define the management operations that can be invoked on nodes or relationships, e.g., creating or deleting a node. TOSCA models are in YAML or XML.

---
[1]https://www.w3.org/Submission/SWRL/

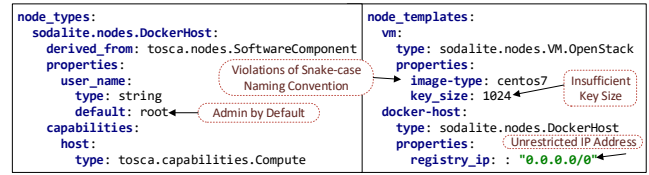

**Figure 1.** Snippets of TOSCA Files Describing a Node Type and an Node Instance, Annotated with Smells

Figure 1 shows TOSCA files describing a node type and a node template. The node type *sodalite.nodes.DockerHost* defines configuration properties, e.g., *user_name*, and specifies its capability to host a Dockerized component. The node template *docker-host* is an instance of this node type. Figure 1 also illustrates some smells, for instance, insecure coding practices of using admin user as the default user, and violation of a naming convention. Such smells deteriorate the quality of deployment model descriptions, and enable the exploitation of vulnerabilities in the deployed systems [11].

### 2.2 Related Work

In software engineering literature [9–11, 13, 15, 16], rule-based reasoning is a common approach to detecting smells and antipatterns. Among these studies, for object-oriented programs, Moha et. al [9] proposed a rule-based domain-specific language (DSL) that supports specification of smells, and automatic generation of detection algorithms. In [10], they have extended their rule-based approach for identifying the antipatterns in service-based systems. The rule-based techniques have been also applied to detect defects in infrastructural code scripts such as Puppet and Chef scripts, e.g., security smells in Puppet [11], and implementation and design smells in Puppet [15] and Chef [13].

Several studies have applied semantic technologies for definition and detection of patterns and antipatterns [3, 12, 14]. Settas et al. [14] modeled the antipatterns in software projects with ontologies, and used a production rule engine to implement detection rules. Inspired by that study, Brabra et al. [3] employed similar semantic technologies to detect antipatterns in cloud service APIs, and to recommend resolutions. Rekiket et al. [12] developed an ontology to represent cloud service offerings, and used common patterns and antipatterns to validate the proposed ontology. They also defined cloud service antipatterns such as invalid VM types and invalid service provider descriptions. Their antipattern detection algorithms employ SPARQL queries.

In this paper, we propose a semantic rule-based approach to detect the smells and antipatterns in descriptions of the deployment models of the cloud applications, for example, smells in TOSCA blueprints (see Figure1). Compared to existing approaches, our framework facilitates the generation of RDF knowledge graphs to capture TOSCA-based deployment models following the conceptual model of DnS. The aim is to map TOSCA to self-contained, independent and reusable knowledge components, amenable to analysis and validation using Semantic Web standards, such as SPARQL.

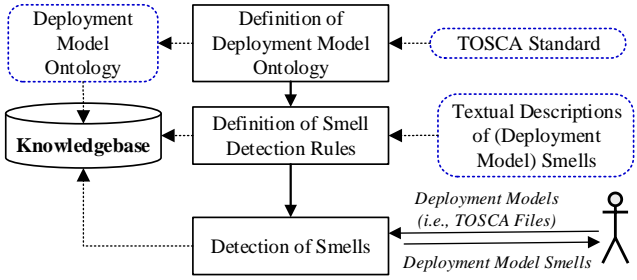

**Figure 2.** An Overview of our Approach

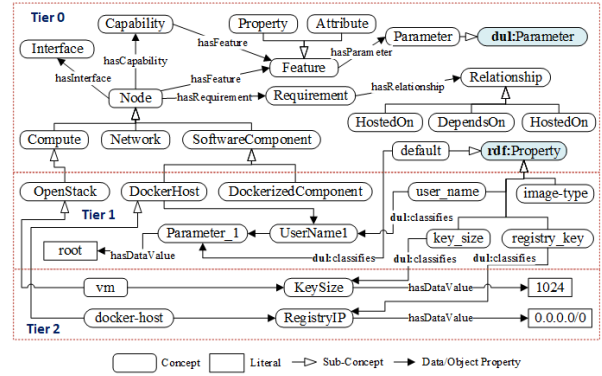

**Figure 3.** Excerpt from Deployment Model Ontology

## 3   Approach

Figure 2 shows the high-level architecture and workflow of our approach to detect the occurrences of smells in deployment model descriptions. More specifically:

- **Definition of Deployment Model Ontology.** To allow a common, extensible and formal standardised model to describe deployment models of cloud applications, we create a deployment model ontology following the DnS (Description and Situation) ontology design pattern, extracting the most important and relevant concepts from the TOSCA standard.
- **Definition of Smells Detection Rules.** After defining the required semantic models, we define the semantic rules in SPARQL to detect the smells in deployment models. The deployment model ontology and detection rules form the knowledge base.
- **Detection of Smells.** Once a developer codifies the deployment topology of an application using TOSCA, he/she can check the occurrence of smells in the created TOSCA file by providing it as inputs to our framework. First, the TOSCA file is translated to an instance of the deployment model ontology. Second, the smell detection rules are applied to detect deployment model-level smells. If a smell is detected, the details of the smell are returned to the developer.

In the following sections, we describe the ontologies and the knowledge-driven detection of smells.

### 3.1   Deployment Model Ontology

For interoperable description of application and infrastructure cloud services, we develop our deployment model ontology based on TOSCA standard. Figure 3 provides an excerpt from the ontology. Due to limited space, we only include key concepts. The complete models are available online (see Section 4). To manage the complexity of defining deployments and to clearly separate modeling roles (e.g., cloud resource expert and application expert), we divide our semantic models into three tiers (aligned with the TOSCA language design).

As shown in Figure 3, Tier-0 captures the key concepts required to describe an application deployment, based on the meta-model of the TOSCA language. *Node* and *Relationship* model the semantics of TOSCA *NodeType* and *Relationship- Type* (see Section 2), which include their capabilities, require- ments, interface, attributes, and properties. The ontology also models the specific types of nodes (e.g., *Compute* and *Network*) and relationships (e.g., *HostedOn* and *DependsOn*)
defined by the TOSCA standard.

Tier-1 defines reusable deployment components, and maps to custom node types in TOSCA. Cloud resource experts can define new node types as necessary, for example, a virtual machine and a Docker container engine. Each such node can have custom properties, capabilities, and interfaces. In our example, the node *DockerHost* has a property of type *user_name* and with the default value as 'root'.

Tier-2 defines the deployment model of an application reusing components, and maps to TOSCA node and relationship templates. In our example, the components *vm* and *docker-host* are nodes of a deployment topology, and instances of the node types *OpenStack* and *DockerHost*. They also instantiate properties of their node types, for example, *registry_ip* as '0.0.0.0/0' address and *key_size* as 1045.

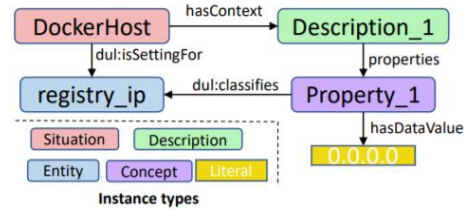

**Figure 4.** Example Instantiation of DnS.

To promote reusability among ontologies at different tiers, we model them as instantiations of the Descriptions and Situations ontology pattern from DOLCE+DnS Ultralite (DUL) [5]. A *Situation* could be a node type, which has a descriptive context (*hasContext*), namely *Description*, which, in turn, describes the concepts of the situation. *Concepts* could be a property, requirement, capability, or other TOSCA concepts, and have zero or more parameters (*hasParameter*). For in- stance, a parameter can have an IP address or a username as a value. Finally, the concepts classify the entities of the situation, which are container classes to represent, for instance, properties, capabilities, and node templates. Figure 4 shows an example instantiation.

### 3.2 Smell Detection Rules

Following software smell detection literature [3,9–11,15,16], we define the deployment model smells as the violations of the best practices or use of the bad practices in designing and codifying deployment models. We map the smells reported in the relevant literature to a deployment model. We first identified the model elements and their properties to detect the smells by amazing the textual definitions of the smells. We next defined the semantic rules needed to detect smells, utilizing the identified concepts and proprieties. As proof of concept, in this paper, we consider 10 smells, and primarily use SPARQL queries for specifying detection rules.

Table 1 shows the (abstract) rules to detect deployment model smells. In the rules, the term *x* represents a property or an attribute of nodes and relationships at both Tier-1 and Tier-2 models. At the Tier-1 model, a property or an attribute can define its default value. For the smell *Suspicious comment*, the term *x* can also be any element in the deployment model. The rules define their logic as expressions of helper func- tions, for example, *isUser()* and *isAdmin()* in the rule *Admin by default*. These functions primarily match string patterns using regular expressions or regular string functions. For example, the function *isUser()* matches the term 'user' to a suffix or prefix of a string. The function *isAdmin()* checks if the property value is either 'admin' or 'root'. These func- tions are based on similar functions reported in the relevant literature [11,15].

```
1  select distinct ?property ?propertyDef
2  where {
3      ?property DUL:classifies ?propertyDef.
4      FILTER(regex(str(?propertyDef),"user(.+?)|(.+?)?user","i")).
5      optional { # node type definitions - tier1
6          ?property DUL:hasParameter ?p .
7          ?p DUL:classifies tosca:default .
8          ?p tosca:hasDataValue ?value.
9      }.
10     optional { # node template definitions - tier0
11         ?property tosca:hasDataValue ?value.
12     }.
13     FILTER (bound(?value)).
14     FILTER (str(?value) IN ('admin', 'root'))
15 }
```

**Figure 5.** Part of AdminByDefault SPARQL Query

Figure 5 shows an excerpt from the SPARQL query for detecting *Admin by default* smell. The variable *proOrAttr* represents a property or attribute of a deployment model element. Line 4 implements the function *isUser* using a regex matching. Lines 5-9 retrieve the default value for a property or attribute. Line 14 realizes the function *isAdmin* using the *IN* operator. The SPARQL queries for the other smells are available online (see the next section).

## 4 Prototype and Evaluation

### 4.1 Prototype

Figure 6 shows the architecture of the prototype implementation. We implement our knowledge base with GraphDB (graphdb.ontotext.com), and use Eclipse RDF4J for manipulating and querying it. Following a service-oriented architecture, we exposed the capabilities of knowledge base and defect predictor as RESTful services using JAX-RS (Java API for RESTful Web Services) standard. We also developed a simple web-based user interface for the defect predictor. All services and UI are web applications that are deployed in Apache Tomcat. The prototype is at the repositories 'semantic-models', 'semantic-reasoner', 'defect-prediction' of our project GitHub (github.com/SODALITE-EU).

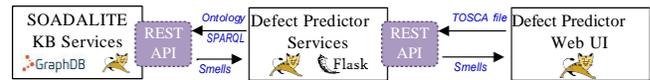

**Figure 6.** Prototype Implementation

By using the REST API of the defect predictor, a user can provide a TOSCA file describing a deployment model. The defect predictor parses the TOSCA file using Open TOSCA Parser (github.com/openstack/tosca-parser), and translates it into the corresponding ontological representation (Tier-1 and Tier-2 models) and stores in the knowledgebase. Then, it executes the SPQRL queries over the stored model to detect smells, and returns a report containing the detected smells back to the user. A demo of the defect/smell predictor is available at *youtube.com/watch?v=IThr5vlleTI*.

### 4.2 Case Studies

We evaluated our defect predictor with three industrial case studies of our European project SODALITE (sodalite.eu), namely clinical trials, vehicle IoT, and Snow. Clinical trials case study focuses the development of a simulation process chain supporting in-silico clinical trials of bone-implant-systems. Vehicle IoT case study deploys a distributed system for processing vehicular data over Cloud and Edge environ- ments. Snow case study deploys a workflow that processes snow images from multiple data sources to derive information on mountain snow coverage.

For each use case, we developed the TOSCA files and the semantic models. We created the buggy version of each TOSCA file by adding 10 smells that we presented in this paper. Then, we validated each buggy TOSCA file with our defect predictor, and analyze the returned smell reports to verify that each smell can be detected successfully. The case study resources are also in our GitHub repositories (see the previous section).

Table 1. Smells, their Descriptions, and the Abstract Detection Rules

| Smell | Smell Description | Abstract Detection Rule |
|---|---|---|
| Admin by default | Default users are administrative users. | $isUser(x.name) \wedge isAdmin(x.name)$ |
| Empty password | A password as a zero-length string. | $isPassword(x.name) \wedge (isEmpty(x.value) \vee isEmpty(x.defaultValue))$ |
| Hard-coded secret | Secrets such as usernames and passwords are hardcoded. | $(isPassword(x.name) \vee isUser(x.name) \vee isSecKey(x.name)) \wedge ((\sim isEmpty(x.value) \wedge \sim isVariable(x.value)) \vee \sim isEmpty(x.defaultValue))$ |
| Suspicious comment | A comment includes the information indicating secrets and buggy implementations. | $hasComment(x) \wedge isSuspicious(x.comment)$ |
| Unrestricted IP address | Using "0.0.0.0" or "::" as binding IP addresses of servers | $isIP(x.name) \wedge (isInvalidBind(x.value) \vee isInvalidBind(x.defaultValue))$ |
| Insecure communication | Using insecure communication protocols, instead of their secure counterparts | $(isURL(x.value) \wedge isInsecure(x.value)) \vee (isURL(x.defaultValue) \wedge isInsecure(x.defaultValue))$ |
| Weak crypto. algo. | Use of weak cryptography algorithms such as MD5 and SHA1 | $hasWeakAlgo(x.value) \vee hasWeakAlgo(x.defaultValue)$ |
| Insufficient key Size | The size of a key used by an encryption algorithm is less than the recommended key size, e.g., 2048 bits for RSA. | $isCryptoKeySize(x.name) \wedge (hasInsufficientKeySize(x.value) \vee hasInsufficientKeySize(x.defaultValue))$ |
| Inconsistent naming convention | The conventions used for naming nodes, properties, attributes, etc., are inconsistent. | $(case=='CamelCase' \rightarrow isCamelCase(x.name)) \vee (case ==' SnakeCase' \rightarrow isSnakeCase(x.name)) \vee (case ==' DashCase' \rightarrow isDashCase(x.name))$ |
| Invalid port ranges | TCP port values are not within the range from 0 to 65535. | $isPort(x.name) \wedge (outOfRange(x.value) \vee outOfRange(x.defaultValue))$ |

## 5  Conclusion and Future Work

In this paper, we have presented an approach that can formally model a cloud application deployment model with ontologies, and detect the smells in the model with ontological reasoning. To show the feasibility of our approach, we developed the support for detecting 10 different smells, and evaluated it with three industrial case studies.

To explain detected smells and recommend fixes, we are currently extending our semantic models to specify smells, their causes, and their fixes. The rule-base is being refined and extended to cover all smells identified by a systematic literature review on infrastructure code smells. We plan to build a unified framework to detect smells across heterogeneous deployment and infrastructure code specifications by utilizing semantic Web techniques such as ontology alignment and query rewriting.

## Acknowledgments

This project has received funding from the European Union's Horizon 2020 research and innovation programme under grant agreement No 825480 (SODALITE project).